\documentclass[conference]{IEEEtran}
\IEEEoverridecommandlockouts
\usepackage{cite}
\usepackage{amsmath,amssymb,amsfonts}
\usepackage{algorithm}
\usepackage{algorithmic}
\usepackage{graphicx}
\usepackage{textcomp}
\usepackage{xcolor}

\usepackage{booktabs}
\usepackage{multirow}
\usepackage{array}
\usepackage{adjustbox}

\usepackage{url}
\usepackage{hyperref}
\hypersetup{breaklinks=true}

\usepackage{tabularx}
\usepackage{threeparttable}
\usepackage{makecell}



\def\BibTeX{{\rm B\kern-.05em{\sc i\kern-.025em b}\kern-.08em
    T\kern-.1667em\lower.7ex\hbox{E}\kern-.125emX}}
\begin{document}

\title{Microarchitectural Co-Optimization for Sustained Throughput of RISC-V Multi-Lane Chaining Vector Processors}

\author{
	\IEEEauthorblockN{Weiying Wang and Zhiwei Zhang}
	\IEEEauthorblockA{
		Institute of Automation, Chinese Academy of Sciences\\
		School of Artificial Intelligence, University of Chinese Academy of Sciences\\
		Beijing 100049, China\\
		Email: \{wangweiying2021, zhiwei.zhang\}@ia.ac.cn
	}
}

\maketitle

\begin{abstract}
Modern RISC vector processors rely on multi-lane parallelism and chaining to achieve high sustained throughput, yet practical execution often deviates from the ideal reference due to microarchitectural inefficiencies. This work targets the open-source RVV processor Ara and analyzes its sustained-throughput loss under a fixed hardware configuration. We first establish an ideal multi-lane chaining model that decomposes ideal execution into prologue startup, steady-state progression, and tail drain, and uses this reference to characterize real-execution deviations. Based on this model, we attribute Ara's bottlenecks to three critical paths: memory-side data supply and transaction progression, dependence-and-issue control, and operand delivery and result propagation. To address these bottlenecks, we propose coordinated optimizations, including a descriptor-driven memory front end with next-VL prefetch, early read-dependence release with dynamic local issue control, and multi-source forwarding with dual-source operand queues. Experimental results show that, without increasing raw memory bandwidth or changing the main processor configuration, Ara-Opt achieves a geometric-mean speedup of $1.33\times$ over baseline Ara. Under roofline-based normalization, the geometric-mean gap-closed ratio reaches 12.2\%. In particular, scal, axpy, ger, and gemm achieve speedups of approximately $2.41\times$, $1.60\times$, $1.52\times$, and $1.42\times$, with corresponding gap-closed ratios of 93.7\%, 88.9\%, 78.3\%, and 59.3\%, respectively. These results show that the proposed optimizations recover lost sustained throughput under essentially unchanged hardware resources and move regular streaming and high-throughput workloads closer to the roofline-based performance bound.
\end{abstract}

\begin{IEEEkeywords}
RISC-V Vector Extension, Sustained Throughput, Multi-Lane Chaining, Microarchitectural Co-Optimization
\end{IEEEkeywords}

\section{Introduction}

As the scale and computational complexity of large models continue to grow~\cite{kaplan2020scalinglawsneurallanguage}, computational throughput and data-supply capability have become key factors limiting overall system efficiency~\cite{ivanov2021datamovementneedcase}. Such models are typically composed of attention, feed-forward, and normalization modules, whose underlying computations can be largely reduced to regular data-parallel operations such as matrix multiplications, element-wise vector operations, and reductions. Therefore, continuously exploiting data-level parallelism under fixed hardware resource constraints has become an important problem for optimizing large-model training and inference.

Vector processors execute multiple data elements with a single instruction, improving throughput while reducing control overhead, and are an important architectural substrate for regular data-parallel computation. Since classic vector machines such as the CRAY-1~\cite{russell1978cray}, vector-register organizations and chaining mechanisms have become fundamental techniques for achieving high sustained throughput in long-vector processors. Although fixed-width SIMD carried vector processing concepts into general-purpose processors, its programming model is tightly coupled to implementation width and therefore limited in scalability. By contrast, the RISC-V Vector Extension (RVV) 1.0~\cite{rvv_spec_1.0}, with its vector-length-agnostic design, provides a unified programming model across implementation scales and establishes an ISA foundation for modern long-vector processors.

Modern RISC vector processors achieve high sustained throughput through the synergy of multi-lane parallelism and chaining: the former determines the upper bound of steady-state execution throughput, while the latter enables dependent vector instructions to overlap as soon as the first results become available. Ideally, if all lanes remain continuously active, the memory system supplies data at a stable rate, dependences are released in a timely manner, and operand delivery incurs no additional conflicts, execution should approach the ideal reference enabled by the available hardware resources. In practice, however, sustained throughput often falls significantly short of this ideal state. This indicates that, beyond peak compute capability and raw memory bandwidth, fine-grained microarchitectural inefficiencies have become a major source of sustained-throughput loss.

In this work, we use the open-source RVV processor Ara as the target platform, analyze its sustained-throughput loss under a fixed hardware configuration, and optimize the design accordingly. We first establish an ideal multi-lane chaining model that decomposes ideal execution into prologue startup, steady-state progression, and tail drain, and uses this reference to characterize real-execution deviations. We then analyze Ara's major bottlenecks along three paths: memory-side data supply, dependence-and-issue control, and operand delivery. Based on this analysis, we propose corresponding coordinated microarchitectural optimizations. Finally, we use a roofline model to evaluate how closely the optimized implementation approaches the performance bound.

The main contributions of this paper are as follows:
\begin{itemize}
    \item We establish an ideal multi-lane chaining model that characterizes the ideal overlapped execution of dependent vector instruction chains under a fixed hardware configuration, and decomposes sustained-throughput loss into additional prologue delay, increased effective steady-state initiation interval, and additional tail overhead.

    \item We analyze Ara's sustained-throughput loss from the perspective of critical execution paths, and attribute the major bottlenecks to discontinuous data supply and inefficient transaction progression on the memory-side path, conservative blocking in dependence-and-issue control, and access conflicts and result-propagation overhead on the operand-delivery path.

    \item We propose coordinated microarchitectural optimizations for the three critical paths, including a descriptor-driven memory front end with next-VL prefetch, early read-dependence release with dynamic local issue control, and multi-source forwarding with dual-source operand queues.

    \item We validate the proposed optimizations in an RTL implementation and show that they improve sustained vector throughput under the same main processor configuration and raw memory bandwidth, moving the performance points closer to the roofline-based performance bound.
    \end{itemize}

\section{Background}

This section summarizes vector architectures, RVV, and the Ara organization, and establishes the ideal multi-lane chaining model used in the subsequent bottleneck analysis.

\subsection{Vector Architecture and RVV}

Vector architectures have evolved from early streaming-style designs to modern scalable vector ISAs. Early implementations, such as the TI ASC and CDC STAR-100, primarily adopted a memory-memory organization and relied on deep pipelines and streaming memory accesses to process large-scale array data~\cite{watson1972ti, STAR-100}. Subsequently, classic vector machines represented by the CRAY-1 established the vector-register paradigm, using low startup overhead and chaining to enable dependent vector instructions to overlap in execution~\cite{russell1978cray}. Later vector supercomputers extended this direction and shaped the long-vector execution model for high sustained throughput~\cite{august1989cray,kawabe1988hitachi,watanabe1987architecture}.

In general-purpose processors, vector concepts were mainly carried forward through fixed-width packed-SIMD extensions, whose programming model is coupled to implementation width and provides limited support for complex memory accesses, tail handling, and long-vector semantics~\cite{intel_sse4_programming_reference,lomont2011introduction}. In contrast, Arm SVE and RISC-V RVV re-emphasize scalable vector lengths and richer vector semantics~\cite{stephens2017arm,rvv_spec_1.0}. RVV 1.0 provides a vector-length-agnostic programming model through \texttt{vl}, \texttt{vtype}, SEW, and LMUL, and supports complex memory accesses, mask/tail handling, reductions, and \texttt{vstart}, providing a unified ISA foundation for modern long-vector processors.

\subsection{Ara Architecture}

Ara is an open-source RVV 1.0-compliant vector processing unit tightly coupled with the CVA6 scalar core~\cite{perotti2022new,perotti2024ara2}. It uses a scalable lane-based organization for vector-level parallel execution. At the top level, Ara consists of the dispatcher, sequencer, lanes, VLSU, SLDU, and MASKU, which handle instruction dispatch and CSR management, dependence and issue control, lane-local execution, vector memory access, slide/permutation operations, and mask processing, respectively. Each lane integrates a local VRF slice and functional units, providing sustained data-parallel throughput over long-vector data streams.

Ara's scalability mainly comes from parameterized configurations of the lane count, VLEN, data types, and related architectural extensions. For the sustained-throughput behavior considered in this work, Ara's performance potential depends on multi-lane utilization and chaining effectiveness. Multi-lane execution provides steady-state throughput, while chaining overlaps dependent vector instructions once early predecessor results become available. These characteristics motivate the ideal execution reference developed next and the subsequent analysis of sustained-throughput loss in the actual implementation.

\subsection{Ideal Multi-Lane Chaining Model}

To analyze sustained-throughput loss in Ara under a fixed hardware configuration, we establish an ideal multi-lane chaining model as a microarchitectural reference. The model keeps the lane count, functional units, vector-register resources, and raw memory bandwidth unchanged, while assuming that implementation-induced inefficiencies are eliminated: stable memory data supply, dependence release at the earliest semantically valid point, operand delivery without additional structural conflicts, and chaining as soon as the first predecessor results become available.

\begin{figure}[t]
\centering
\includegraphics[width=0.48\textwidth]{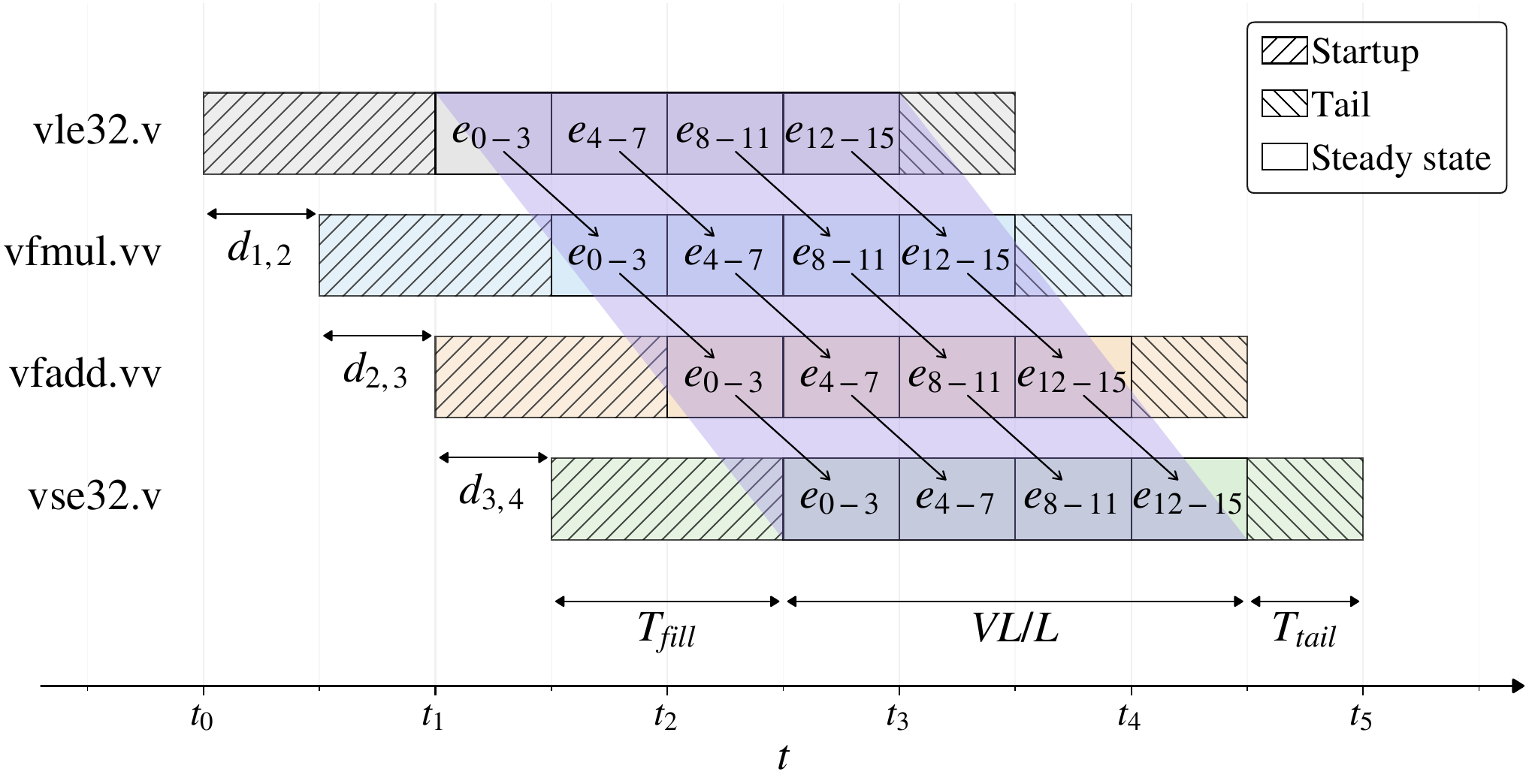}
\caption{Execution timeline and total-execution-time decomposition of a dependent vector instruction chain under ideal multi-lane chaining. The shaded region denotes the peak-throughput interval, where the instruction chain reaches steady-state overlap, and multiple vector instructions execute concurrently on different element groups.}
\label{fig:ideal_model}
\end{figure}

Fig.~\ref{fig:ideal_model} illustrates the execution timeline of a dependent vector instruction chain,
\texttt{vle32.v} $\rightarrow$ \texttt{vfmul.vv} $\rightarrow$ \texttt{vfadd.vv} $\rightarrow$ \texttt{vse32.v},
under ideal multi-lane chaining. Let $d_{i,i+1}$ denote the minimum startup-propagation delay between two adjacent dependent instructions, i.e., the unavoidable delay before the first results of instruction $i$ can be consumed by instruction $i+1$. Let $T_{\mathrm{fill}}$ denote the additional fill time from the startup of the last stage until the entire chain enters stable overlapped execution. The ideal prologue time is therefore
\begin{equation}
p_N
=
\sum_{i=1}^{N-1} d_{i,i+1}
+
T_{\mathrm{fill}} .
\end{equation}

For vector length $VL$ and lane count $L$, the ideal steady-state phase advances one element group per cycle. Thus, the ideal steady-state execution time is
\begin{equation}
T_{\mathrm{steady}}^{\mathrm{ideal}}
=
\left\lceil \frac{VL}{L} \right\rceil .
\end{equation}

The ideal total execution time of the dependent instruction chain is
\begin{equation}
T_{\mathrm{ideal}}
=
p_N
+
T_{\mathrm{steady}}^{\mathrm{ideal}}
+
T_{\mathrm{tail}},
\end{equation}
where $T_{\mathrm{tail}}$ denotes the tail-drain time. This expression shows that, under ideal chaining, a dependent vector instruction chain is not executed as a serial accumulation of individual instruction latencies. Instead, its execution consists of a one-time prologue, a continuously overlapped steady-state phase, and a one-time tail drain.

We then express real execution as a deviation from the ideal reference:
\begin{equation}
T_{\mathrm{real}}
=
(p_N+\Delta p)
+
T_{\mathrm{steady}}^{\mathrm{ideal}}
\cdot
II_{\mathrm{eff}}
+
(T_{\mathrm{tail}}+\Delta t).
\end{equation}

Here, $\Delta p$ denotes the additional prologue delay beyond ideal startup propagation; $II_{\mathrm{eff}}$ denotes the effective number of cycles required to advance one element group in the steady state; and $\Delta t$ denotes additional tail overhead that cannot be hidden by overlap. In the ideal case, $\Delta p=0$, $II_{\mathrm{eff}}=1$, and $\Delta t=0$. Therefore, Ara's sustained-throughput loss can be written as
\begin{equation}
\Delta T
=
T_{\mathrm{real}}
-
T_{\mathrm{ideal}}
=
\Delta p
+
T_{\mathrm{steady}}^{\mathrm{ideal}}
(II_{\mathrm{eff}}-1)
+
\Delta t .
\end{equation}

For long-vector and regular streaming workloads, $T_{\mathrm{steady}}^{\mathrm{ideal}}$ usually dominates the total execution time. Therefore, even a slight increase in $II_{\mathrm{eff}}$ can accumulate into significant throughput loss during the steady-state phase. The subsequent analysis uses this decomposition to identify the microarchitectural factors that enlarge $\Delta p$, increase $II_{\mathrm{eff}}$, or increase $\Delta t$.

\section{Related Work}

Modern scalable vector processors, especially RVV-based designs and adjacent design spaces, have formed a rich implementation spectrum. The Ara family represents the long-vector, multi-lane RVV direction~\cite{cavalcante2019ara,perotti2022new,perotti2024ara2}, while Saturn, Vitruvius+, and Spatz cover design points such as short-vector RVV, hybrid long-vector coprocessors, and shared-L1 vector units~\cite{zhao2024instruction,minervini2023vitruvius,cavalcante2022spatz}. Hwacha demonstrated the feasibility of a decoupled scalar/vector front end with a vector-fetch organization~\cite{lee2015hwacha}. Prior studies have also explored vector-throughput optimization from local perspectives, including short-vector chaining and scheduling, vector-register renaming or remapping, arithmetic/memory decoupling, dedicated support for reductions, and mechanisms related to bank conflicts, arbitration, and operand buffering~\cite{zhao2024instruction,minervini2023vitruvius,patsidis2020risc,razilov2025conflict}. Overall, these studies mainly focus on implementation organizations, design-space choices, or local mechanisms, rather than providing a unified attribution of sustained-throughput loss across complete execution paths in an existing implementation.

The works most closely related to ours analyze practical vector-processor throughput using ideal references or efficiency-oriented metrics. Ara2 introduces \emph{raw throughput ideality} and the ideal dispatcher to quantify how far the vector backend deviates from an ideal execution state~\cite{perotti2024ara2}. TROOP analyzes throughput loss in low-operational-intensity workloads from a roofline perspective~\cite{purayil2025troop}. RISCV2 uses \emph{elements per cycle} (EPC) to reflect changes in effective parallelism brought by mechanism-level optimizations~\cite{patsidis2020risc}. These studies show that absolute performance alone is insufficient for explaining sustained-throughput behavior; appropriate references or efficiency metrics are needed to relate practical execution to ideal capability. Unlike prior work that proposes new design points or optimizes isolated local mechanisms, this work targets an existing Ara implementation and analyzes why it falls short of the ideal sustained-throughput reference under a fixed processor configuration and raw memory-bandwidth constraints. We further provide an execution-path-level attribution of the loss and evaluate the recoverable throughput through coordinated optimizations.

\section{Sustained-Throughput Bottleneck Analysis}

This section analyzes Ara's sustained-throughput loss based on the ideal multi-lane chaining model, focusing on the memory-side data-supply path, the dependence-and-issue control path, and the operand-delivery path.

\subsection{Memory-Side Data Supply and Transaction Inefficiency}

The first class of bottlenecks arises along the memory-side path. Ideal steady-state execution requires load data to return continuously at a rate matched to the consumption rate of the lane backend, thereby maintaining \(II_{\mathrm{eff}}=1\) and keeping load, compute, and store stages overlapped. In Ara, however, load data delivery is primarily demand-driven: data are passively returned after requests have been issued. When latency in the external memory hierarchy is not hidden in advance, the backend may stall due to discontinuous input data even with sufficient parallelism, reducing steady-state progression efficiency and appearing in the model as \(II_{\mathrm{eff}}>1\).

This problem is further amplified by inefficient transaction organization in the memory front end. In the current implementation, memory-instruction reception, address expansion, transaction generation, and transaction issuance are tightly coupled. A single vector memory instruction can occupy front-end resources throughout address expansion, while bus-handshake stalls can propagate back to transaction generation and address expansion. In addition, read and write requests are not sufficiently separated on the issue path, allowing different transaction types to interfere with each other. As a result, the memory-side path cannot sustain a stable transaction stream, and data-supply gaps are exposed in the steady-state phase, forming the memory-side component of \(T_{\mathrm{steady}}^{\mathrm{ideal}}(II_{\mathrm{eff}}-1)\).

\subsection{Conservative Dependence and Issue Control}

The second bottleneck comes from conservative dependence management and issue control, which delays the startup of dependent instructions beyond the ideal chaining point. Ideal chaining requires dependences to be released at the earliest semantically safe point, allowing a successor instruction to start as soon as the first predecessor results become available and dataflow constraints are satisfied. Therefore, the actual startup interval between adjacent dependent instructions should approach the minimum propagation delay \(d_{i,i+1}\). In Ara, however, dependence release is conservative: the associated protection is often retained until the entire instruction completes, rather than being released after source operands have been read and the read-related dependence has been resolved. This introduces additional control delay beyond \(d_{i,i+1}\), thereby increasing \(\Delta p\).

Lane-local issue control can also introduce steady-state bubbles. Some operand requests may remain valid even though they have already satisfied downstream handshake conditions and can be consumed in the current cycle. If the control logic blocks subsequent issues solely based on the valid status, the occupancy that could have been released is still treated as an effective blocking source. Such short-cycle stalls recur during long-vector execution and accumulate into a control-side component of \(II_{\mathrm{eff}}>1\).

\subsection{Operand-Delivery Inefficiency}

The third bottleneck lies in the operand-delivery path, where structural conflicts and indirect result propagation can serialize otherwise overlapped execution. Ideal execution requires operands to reach consumers after the minimum propagation delay without additional serialization from structural conflicts. Therefore, operand delivery should neither increase \(\Delta p\) nor break the steady-state condition \(II_{\mathrm{eff}}=1\). In Ara, although the VRF is partitioned across lanes, intra-lane source-operand reads, result write-backs, and data exchanges among functional units still contend at the bank, port, and arbitration levels. When multiple requests map to the same structural resource, accesses are serialized, disrupting continuous element-group progression and increasing \(II_{\mathrm{eff}}\).

Meanwhile, results produced by predecessor functional units still need to be written back to the VRF and then reread by successor consumers, stretching operand propagation into an indirect produce--write-back--reread path. This path increases the actual propagation delay between adjacent dependent instructions and enlarges \(\Delta p\); it also introduces additional VRF read/write traffic, aggravating structural conflicts and increasing steady-state bubbles. Therefore, operand-delivery inefficiency affects both \(\Delta p\) and \(II_{\mathrm{eff}}\), weakening the ideal overlapped execution enabled by multi-lane chaining.

\section{Microarchitectural Co-Optimization}

\begin{figure*}[t]
\centering
\includegraphics[width=0.82\textwidth]{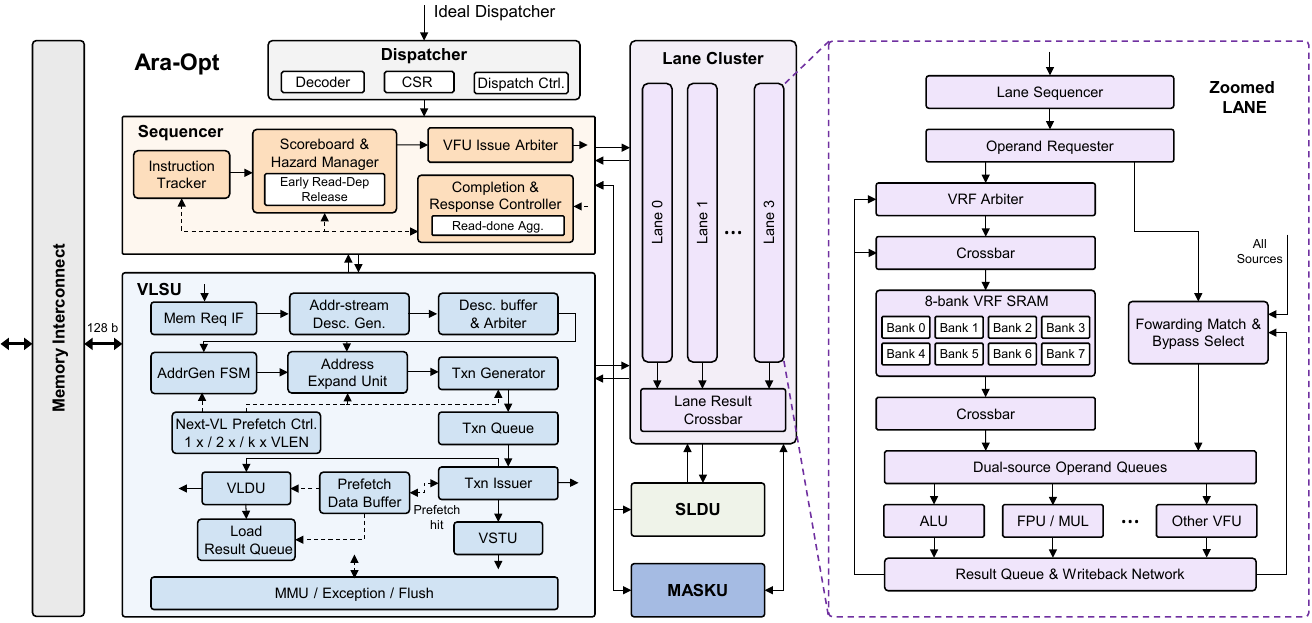}
\caption{Microarchitectural overview of Ara-Opt.}
\label{fig:main}
\end{figure*}

To address the three bottleneck classes identified above, we co-optimize Ara along three execution paths: the memory front end, dependence management and issue control, and operand delivery. Fig.~\ref{fig:main} shows the overall Ara-Opt microarchitecture.

\subsection{Decoupled Memory Front End and Next-VL Prefetch}

To address discontinuous memory-side data supply and inefficient transaction progression, Ara-Opt first decouples the memory front end in the VLSU. After a vector memory request enters the front end, the address-stream descriptor generator encapsulates the access characteristics of the entire instruction, including the starting address, remaining access length, stride, element width, and access type. The descriptor is then passed through the descriptor buffer and arbiter to the address-generation control unit for centralized scheduling. The address-generation control unit uses a finite-state machine to drive the address expansion unit, generating addresses for unit-stride, strided, and indexed access modes, and entering a waiting state when address translation is required. The transaction generator then converts the expanded addresses into bus transactions, the transaction queue buffers pending transactions, and the transaction issuer continuously sends requests to the bus interface. By introducing buffering boundaries among instruction reception, address expansion, transaction generation, and bus issuance, the memory front end reduces the impact of backend backpressure and can continuously issue transactions when the queue is not full and the bus is ready, thereby reducing memory-side bubbles caused by unstable transaction streams.

On top of this decoupled front end, we introduce next-VL prefetch to reduce the load path's direct dependence on immediate demand returns. For sequential and predictable unit-stride accesses, the next-VL prefetch control unit derives prefetch addresses from the current access stream and the subsequent vector-length interval, and proactively organizes prefetch transactions for the next VL interval. Prefetch requests and regular demand requests are distinguished using different AXI IDs, and returned prefetch data are stored in the prefetch data buffer. When a subsequent load hits in the prefetch window, the data can be delivered directly from the prefetch buffer to the VLDU, the load result queue, and the subsequent write-back path, without waiting for another return from the external memory hierarchy. In this way, the load path is transformed from a passive supply scheme that fully depends on demand returns into an active supply mechanism in which demand requests serve current accesses while next-VL prefetch prepares future data in the background, improving the continuity of load data supply.

\subsection{Early Read-Dependence Release and Dynamic Local Issue}

To mitigate conservative blocking in dependence management and issue control, Ara-Opt first introduces early read-dependence release in the scoreboard and hazard manager of the main sequencer. After each lane's operand requester completes the source-operand reads of an in-flight instruction, it returns a read-done status, which is globally aggregated by the read-done aggregator in the completion/response controller. Once the system confirms that all source operands of the instruction have entered the corresponding lane operand queues, the sequencer clears its read occupancy in the read list, instead of waiting for the entire instruction to complete. This mechanism only releases the reader occupancy required for WAR checks early, while the write list used for RAW/WAW dependences remains occupied until the producer instruction completes. Therefore, it preserves true data-dependence ordering while moving read-dependence release from instruction completion to source-operand consumption, allowing successor write instructions to pass hazard checks earlier and reducing false blocking.

Ara-Opt also introduces dynamic release-aware local issue control in the lane sequencer. For an operand request that remains valid but whose downstream module is ready and can consume it in the current cycle, the control logic no longer treats it as a hard blocking source. Accordingly, the lane-local issue condition is extended from a purely static occupancy check to a joint check of occupancy status and current-cycle releasability, so that operand-requester occupancy that can be released in the current cycle does not linger unnecessarily. Global early read-dependence release and local dynamic issue control reduce cross-instruction false-dependence stalls and lane-local blocking, respectively, allowing successor instructions to enter execution closer to the earliest semantically safe point.

\subsection{Forwarding-Enhanced Operand Delivery}

To reduce result-propagation overhead and VRF access conflicts on the operand-delivery path, Ara-Opt introduces multi-source operand forwarding in the operand requester. In the original path, predecessor results typically need to be written back to the VRF and then reread by successor instructions through the VRF arbiter and operand crossbar. This long produce--write-back--reread path increases dependence-propagation latency and aggravates VRF access pressure. Ara-Opt adds forwarding match and bypass select logic, allowing the operand requester to monitor result channels from load, ALU, FPU/MUL, MASKU, SLDU, and other VFUs in addition to normal VRF reads. When the dependence relation, requested register, and element position match the predecessor result, the result bypasses the VRF reread path and is sent directly to the corresponding operand queue, allowing the successor consumer to proceed as soon as the predecessor result becomes visible.

To make forwarding a stable source of benefit, Ara-Opt extends the operand queues into dual-source operand queues that support writes from both the VRF read path and the forwarding network. When a VRF-read result and a forwarded result arrive in the same cycle and sufficient queue space is available, both can be enqueued simultaneously. When space is limited, the control logic selects the accepted input according to the queue state and request status, ensuring that the original VRF-read semantics are not violated. With this dual-source input structure, forwarded results are no longer temporary combinational bypasses but become queueable and bufferable operand sources. Multi-source forwarding and dual-source operand queues shorten result-propagation latency along dependent instruction chains and reduce unnecessary VRF reread requests, thereby lowering VRF bank, port, and arbitration pressure within each lane and moving the overlap between adjacent dependent instructions closer to ideal chaining behavior.

\section{Evaluation}

This section evaluates the effectiveness of Ara-Opt, focusing on sustained-throughput improvement, roofline gap closure, scale sensitivity, performance-gain attribution, and efficiency/design positioning.

\subsection{Experimental Setup}

The evaluation compares baseline Ara against Ara-Opt, which integrates the proposed optimizations. The two designs use the same main architectural configuration and raw memory bandwidth. All experiments are conducted with VLEN=1024, DLEN=256, and a 128-bit AXI interface. Performance results are obtained from cycle-accurate RTL simulation. The simulation uses the Ideal Dispatcher mode, which injects vector instructions into Ara at the maximum feasible issue rate, thereby reducing the impact of the scalar-core front end on the measured performance and focusing the evaluation on the execution efficiency of Ara's vector backend and memory system. Area and power results are derived from synthesis using the TSMC 28nm HPC+ technology library and gate-level switching activity at the tt / 0.9\,V / 25$^\circ$C corner.

To cover different execution characteristics, we use a representative set of vector kernels implemented in hand-optimized assembly. scal and axpy represent regular streaming and load-compute-store overlap. dotp, gemv, and symv cover reduction or accumulation dependences and BLAS2-style mixed memory/compute workloads. gemm, trsm, and syrk represent higher arithmetic intensity and block-level data reuse, while ger complements the set with regular matrix updates and two-dimensional streaming access. spmv and dwt represent irregular or complex memory-access workloads, respectively. For one-dimensional vector kernels, long vector lengths are used to reduce the influence of startup and tail overheads. For matrix kernels, representative problem sizes are selected, and gemm is additionally evaluated with multiple sizes to analyze scaling effects.

\subsection{Performance, Gap Closure, and Sensitivity}

\begin{figure}[t]
\centering
\includegraphics[width=0.48\textwidth]{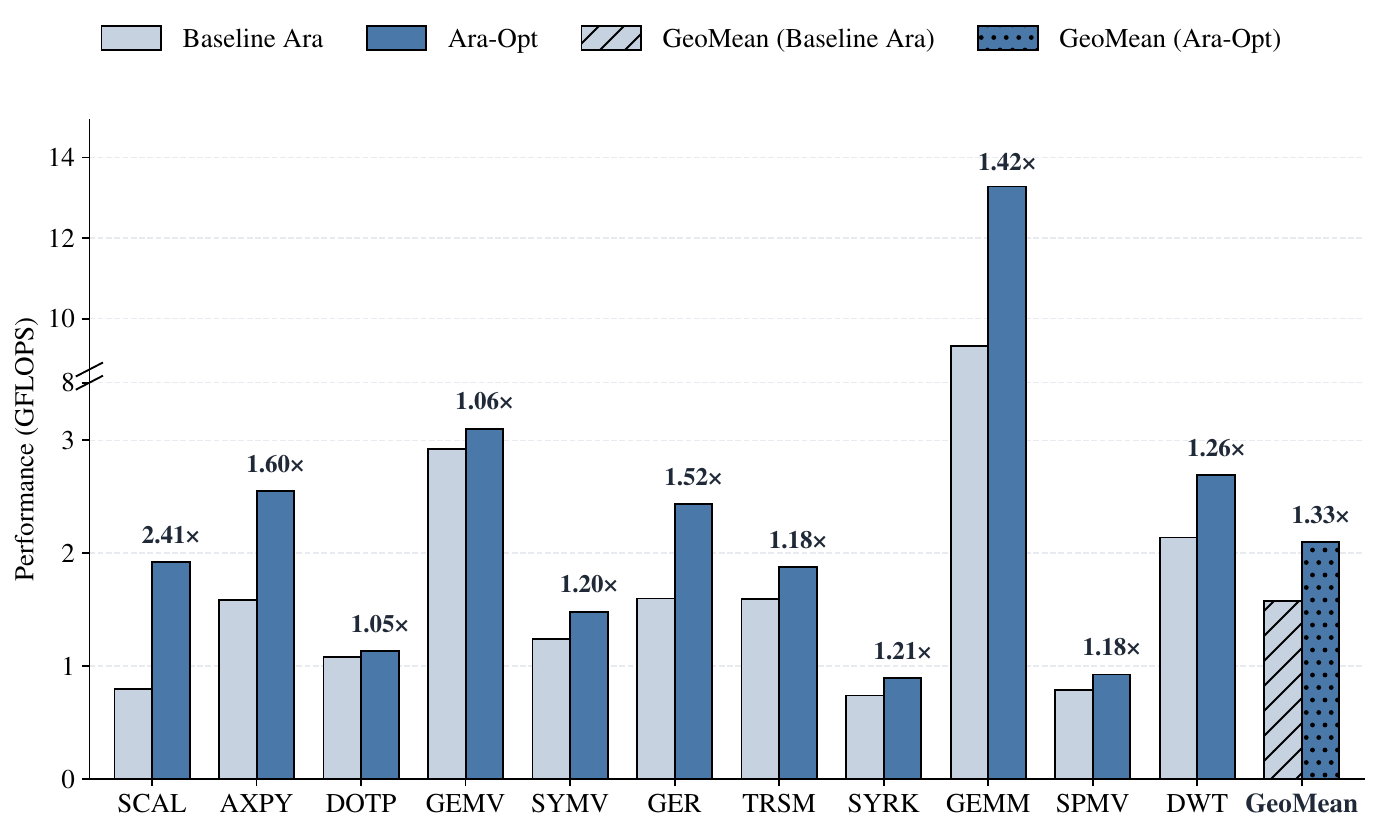}
\caption{Achieved performance of baseline Ara and Ara-Opt across workloads. Speedups are annotated above Ara-Opt bars, with the geometric mean reported over all workloads. Unless otherwise specified, problem sizes are \(N=1024\) for 1-D kernels, \(32\times128\) for gemv, \(32\times32\) for symv, trsm, syrk, and spmv, and \(128\times128\) for ger and default gemm.}
\label{fig:perf}
\end{figure}

Fig.~\ref{fig:perf} compares the achieved performance of baseline Ara and Ara-Opt. Ara-Opt improves all evaluated kernels and achieves a geometric-mean speedup of approximately $1.33\times$, indicating that the proposed optimizations improve realized sustained throughput across different execution modes. The largest gains are observed on scal, axpy, ger, and gemm, with speedups of about $2.41\times$, $1.60\times$, $1.52\times$, and $1.42\times$, respectively. symv, syrk, dwt, trsm, and spmv achieve stable gains of about $1.18\times$--$1.26\times$, whereas dotp and gemv improve more modestly, by about $1.05\times$ and $1.06\times$, respectively.

\begin{figure}[t]
\centering
\includegraphics[width=0.48\textwidth]{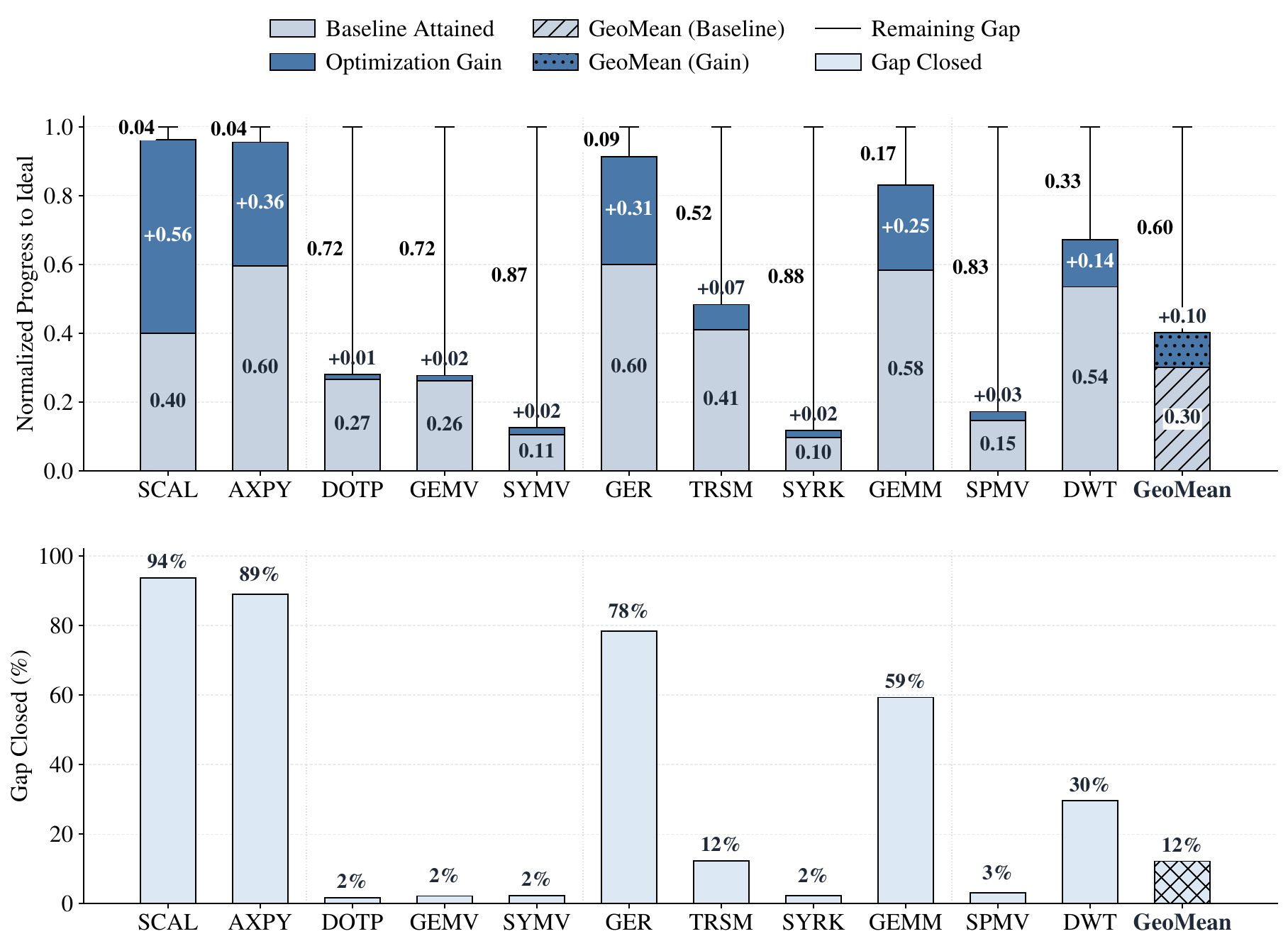}
\caption{Normalized progress toward the roofline-based performance bound. The upper panel shows the baseline-attained fraction, the additional fraction recovered by Ara-Opt, and the remaining gap; the lower panel reports the corresponding gap-closed ratio.}
\label{fig:gap}
\end{figure}

To evaluate how closely the performance gains approach the roofline-based performance bound~\cite{RooflineWilliams}, Fig.~\ref{fig:gap} normalizes the measured performance of each kernel to the roofline upper bound. The peak compute capability is set to \(P_{\mathrm{peak}} = 16\,\mathrm{GFLOPS}\), and the peak memory bandwidth is set to \(BW = 16\,\mathrm{GB/s}\). The ideal performance of each kernel is defined as
\[
P_{\mathrm{ideal}}
=
\min\!\left(P_{\mathrm{peak}},\, BW \times OI\right),
\qquad
OI
=
\frac{\text{kernel\_ops}}{\text{kernel\_bytes}} .
\]
Fig.~\ref{fig:gap} reports the fraction already attained by the baseline, the additional fraction recovered by Ara-Opt, and the remaining gap. Ara-Opt moves most workloads closer to the roofline-based performance bound, although the gap closure remains workload dependent. For scal, axpy, and ger, normalized performance improves from 0.40, 0.60, and 0.60 to 0.96, 0.95, and 0.91, respectively, corresponding to gap-closed ratios of 93.7\%, 88.9\%, and 78.3\%. For gemm, normalized performance improves from 0.58 to 0.83, corresponding to a gap-closed ratio of 59.3\%. By contrast, dotp, gemv, symv, syrk, and spmv exhibit relatively low gap-closed ratios, indicating that their remaining gaps are more strongly constrained by reduction/accumulation serialization, data-reuse patterns, or irregular access behavior. Overall, the geometric mean of normalized performance increases from 0.30 to 0.40, and the geometric-mean gap-closed ratio reaches 12.2\%.

\begin{figure}[t]
\centering
\includegraphics[width=0.48\textwidth]{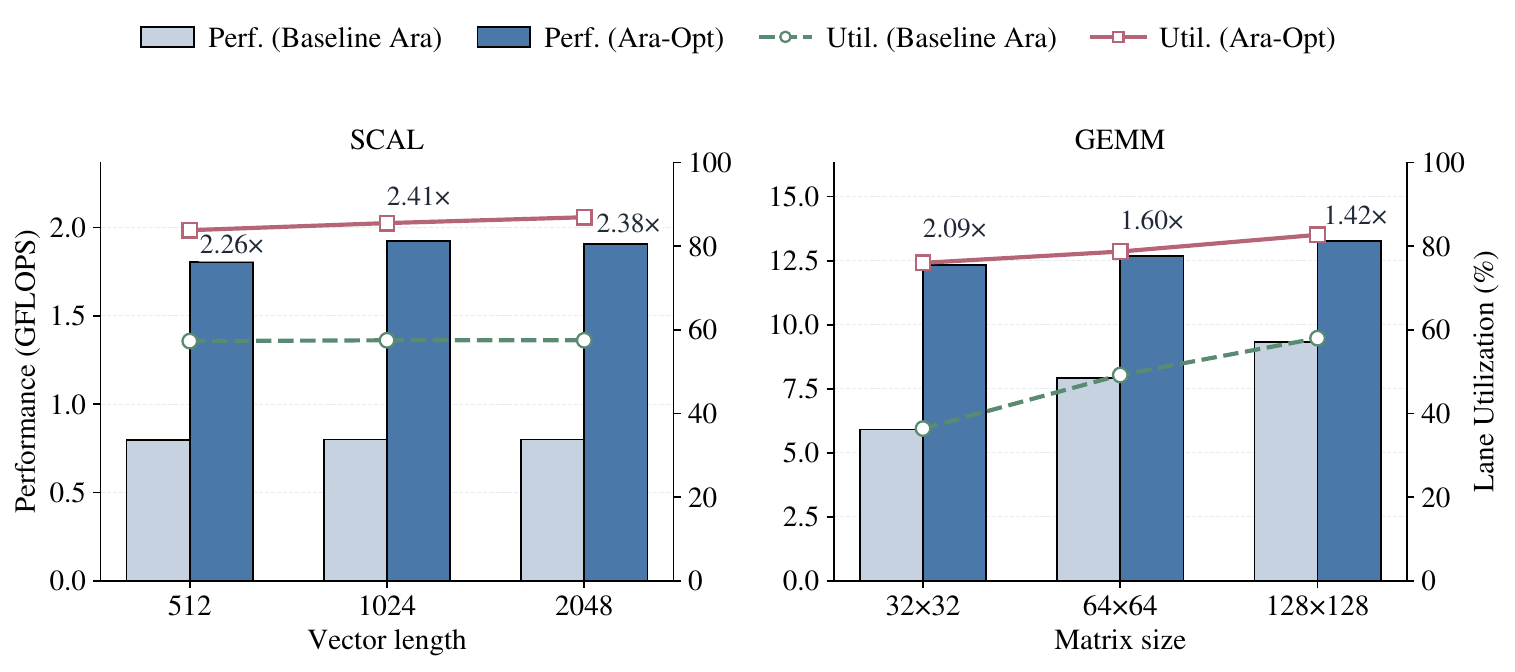}
\caption{Problem-size sensitivity for scal and gemm. Bars show achieved performance for baseline Ara and Ara-Opt, and lines show lane utilization. Ara-Opt maintains stable benefits for scal and improves absolute gemm performance, while relative gemm speedup gradually converges with matrix size.}
\label{fig:sensitivity}
\end{figure}

Fig.~\ref{fig:sensitivity} further analyzes the impact of problem size on optimization benefit. For scal, Ara-Opt maintains stable gains over the baseline across \(N=512\) to \(N=2048\). For gemm, as the matrix size increases, both absolute performance and lane utilization continue to improve, while the relative speedup gradually converges. This indicates that higher data reuse and compute density can amortize part of the microarchitectural inefficiency.

\subsection{Ablation and Attribution of Performance Gains}

To quantify the individual contributions and interactions of the three optimization classes, we perform a \(2^3\) orthogonal ablation study, as shown in Table~\ref{tab:ablation}. The GeoMean in this table is computed only over the selected ablation workloads and is used to summarize mechanism-level contribution trends.

\begin{table}[t]
\centering
\caption{Orthogonal ablation of Ara-Opt optimizations.}
\label{tab:ablation}
\begin{threeparttable}
\setlength{\tabcolsep}{0pt}
\renewcommand{\arraystretch}{1.05}
\begin{tabular*}{0.99\columnwidth}{@{\extracolsep{\fill}}lccccccc}
\toprule
\textbf{Kernel} & \textbf{M} & \textbf{C} & \textbf{O} & \textbf{M+C} & \textbf{M+O} & \textbf{C+O} & \textbf{All} \\
\midrule
scal    & 1.24 & 1.36 & 1.14 & 2.09 & 1.47 & 1.52 & 2.41 \\
axpy    & 1.22 & 1.05 & 1.03 & 1.59 & 1.12 & 1.11 & 1.60 \\
ger     & 1.13 & 1.05 & 1.03 & 1.45 & 1.03 & 1.11 & 1.52 \\
gemm    & 1.26 & 1.05 & 1.10 & 1.41 & 1.29 & 1.12 & 1.42 \\
gemv    & 1.07 & 1.00 & 1.07 & 1.01 & 1.07 & 1.07 & 1.06 \\
dotp    & 1.00 & 1.04 & 1.04 & 1.02 & 1.04 & 1.06 & 1.05 \\
\midrule
GeoMean & 1.15 & 1.09 & 1.07 & 1.38 & 1.16 & 1.16 & 1.45 \\
\bottomrule
\end{tabular*}
\begin{tablenotes}[flushleft]
\footnotesize
\item \emph{Note:} M, C, and O denote memory-side, control-side, and operand-delivery optimizations; values are speedups over baseline Ara.
\end{tablenotes}
\end{threeparttable}
\end{table}

Among the single-class configurations, M provides the largest independent benefit, with a GeoMean speedup of \(1.15\times\), compared with \(1.09\times\) for C and \(1.07\times\) for O. In particular, M improves scal, axpy, ger, and gemm by \(1.24\times\), \(1.22\times\), \(1.13\times\), and \(1.26\times\), respectively. This indicates that memory-side data-supply continuity and transaction progression are important limiting factors for regular streaming, regular matrix-update, and high-throughput workloads. By contrast, the standalone benefits of C and O are more modest, suggesting that control-side blocking and operand-delivery inefficiencies are usually not isolated dominant bottlenecks, but instead interact with data-supply timing and steady-state progression.

The combined configurations further reveal the synergy among the optimization classes. M+C reaches a GeoMean speedup of \(1.38\times\), close to the \(1.45\times\) achieved by the full Ara-Opt configuration, and approaches the All configuration on axpy, ger, and gemm. This shows that stable data supply must be coupled with timely dependence release and local issue control to sustain the steady-state progression enabled by multi-lane chaining. The standalone benefit of O is relatively small, but it reaches \(1.10\times\) on gemm, and M+O further improves gemm to \(1.29\times\). This indicates that operand-delivery optimization still provides complementary benefit under high register-access pressure.

Runtime behavior is consistent with the ablation results. For the workloads with larger gains, Ara-Opt significantly improves lane compute utilization: scal increases from 10.0\% to 24.1\%, axpy from 9.9\% to 15.9\%, ger from 10.0\% to 15.2\%, and gemm from 58.0\% to 82.7\%. For gemm, the VRF bank conflict ratio also drops from 14.0\% to 5.0\%. These results show that the proposed optimizations not only reduce execution time, but also convert more steady-state cycles into useful vector computation.

By contrast, dotp and gemv show limited gains across all configurations, with the All configuration reaching only \(1.05\times\) and \(1.06\times\), respectively. This behavior is consistent with their stronger reduction or accumulation dependences: the recoverable steady-state overlap from improving data supply, issue control, and operand delivery is limited, so alleviating local path inefficiencies does not translate into substantial overall speedup. Overall, the ablation study shows that Ara-Opt is primarily driven by coordinated recovery of memory-side data supply and dependence-and-issue control, while operand-delivery optimization provides additional gains in high-throughput workloads with higher register-access pressure.

\subsection{Efficiency and Design Positioning}
Table~\ref{tab:fixed_resource_eff} summarizes the PPA and efficiency comparison between Ara and Ara-Opt. Ara-Opt achieves a $1.42\times$ throughput improvement on gemm, and improves area efficiency from 3.53~GFLOPS/mm$^2$ to 4.78~GFLOPS/mm$^2$. This comes with a 5.30\% area increase and a 50.86\% power increase, reducing energy efficiency slightly from 65.68~GFLOPS/W to 62.04~GFLOPS/W. These results show that the proposed optimizations improve the effective utilization of existing hardware resources with modest area overhead, while the recovered throughput also leads to higher hardware activity and power cost. Overall, Ara-Opt improves realized throughput and area efficiency mainly by recovering sustained-throughput loss along existing execution paths, rather than by raising the performance bound through hardware scaling or processor reorganization. This suggests that mitigating microarchitectural inefficiencies under a fixed configuration is an effective step before scaling hardware resources.

\begin{table}[t]
\centering
\caption{PPA and efficiency summary of Ara and Ara-Opt.}
\label{tab:fixed_resource_eff}
\scriptsize
\setlength{\tabcolsep}{0pt}
\renewcommand{\arraystretch}{1.08}

\begin{threeparttable}
\begin{tabular*}{0.98\columnwidth}{@{\extracolsep{\fill}}lccc@{}}
\toprule
\textbf{Metric} & \textbf{Ara} & \textbf{Ara-Opt} & \textbf{Relative Change} \\
\midrule
Technology & TSMC 28\,nm & TSMC 28\,nm & -- \\
Frequency (GHz) & 1.00 & 1.00 & -- \\
Lanes / VLEN / DLEN & 4/1024/256 & 4/1024/256 & -- \\
Achieved Perf. (GFLOPS) & 9.32 & 13.28 & 1.42$\times$ \\
Area (mm$^2$) & 2.64 & 2.78 & +5.30\% \\
Power (mW) & 141.89 & 214.05 & +50.86\% \\
Energy Eff. (GFLOPS/W) & 65.68 & 62.04 & 0.94$\times$ \\
Area Eff. (GFLOPS/mm$^2$) & 3.53 & 4.78 & 1.35$\times$ \\
\bottomrule
\end{tabular*}

\begin{tablenotes}[flushleft]
\scriptsize
\item[] \textit{Note:} Lanes/VLEN/DLEN denotes count/bits/bits; performance and power are measured on single-precision $128\times128$ gemm.
\end{tablenotes}
\end{threeparttable}
\end{table}

\section{Conclusions}
This work studies sustained-throughput loss in RVV multi-lane chaining vector processors under fixed resource constraints. Using Ara as the target platform, we establish an ideal multi-lane chaining model and analyze practical deviations along critical execution paths.
The analysis shows that Ara's sustained-throughput loss mainly comes from discontinuous memory-side data supply and inefficient transaction progression, conservative blocking in dependence-and-issue control, and access conflicts and propagation overhead along the operand-delivery and result-propagation path. Ara-Opt addresses these bottlenecks with a descriptor-driven memory front end and next-VL prefetch, early read-dependence release and dynamic local issue control, and multi-source forwarding with dual-source operand queues.

RTL evaluation shows that Ara-Opt achieves a geometric-mean speedup of $1.33\times$ and a geometric-mean gap-closed ratio of 12.2\% without increasing raw memory bandwidth or changing the main processor configuration. The largest gains appear on regular streaming and high-throughput workloads, with speedups ranging from $1.42\times$ to $2.41\times$. These results show that the proposed approach effectively recovers sustained-throughput loss caused by microarchitectural inefficiencies in Ara and moves regular streaming and high-throughput workloads closer to the roofline-based performance bound. For reduction-dominated workloads and complex non-contiguous memory-access scenarios, the current approach remains less effective. Future work will further investigate cross-element dependence scheduling and reduction-result propagation for reduction-dominated workloads, as well as prefetch decision, transaction organization, and memory-access coalescing for complex memory-access scenarios.

\bibliographystyle{IEEEtran}
\bibliography{references}

\end{document}